\documentclass[12pt]{article}

\def\sepand{\rule{14cm}{0pt}\and}

\begin{document}

\title{A new white dwarf constraint on the rate of change of the gravitational constant}

\vspace{0.6cm}

\author{{\sc Marek Biesiada}\\
\sepand
{\sc Beata Malec}\\
{\sl Department of Astrophysics and Cosmology,}\\
{\sl University of Silesia,}\\
{\sl Uniwersytecka 4, 40-007 Katowice, Poland}\\
{mb@imp.sosnowiec.pl;  beata@server.phys.us.edu.pl}\\
}
\date{}
\maketitle \vfill

\begin{abstract}

In this paper we derive a bound on the rate of change of the
gravitational constant $G$ coming from the pulsating white dwarf
G117-B15A. This star is a ZZ Ceti pulsator extensively studied
with astroseismological techniques for last three decades. The
most recent determination of ${\dot P} = (2.3 \pm 1.4) \times
10^{-15} \;\;s s^{-1}$ for the $215.2\;s$ fundamental mode agrees
very well with predictions of the best fit theoretical model. 
The rate of change of the oscillation period can be explained by two 
effects: the cooling (dominant factor) and change of gravitational 
binding energy (residual gravitational contraction). 
Since the white dwarfs are pulsating in 
g-modes whose frequencies are related to the 
Brunt-V$\rm {\ddot a}$is$\rm {\ddot a}$l$\rm {\ddot a}$ 
frequency (explicitly dependent on $G$) 
observational determination of the change of  
the period (more precisely the difference between observed and 
calculated $\dot P$) 
can be
used to set the upper bound on the rate of change of $G$. In the
light of the current data concerning G117-B15A we derive the
following bound: $|{\frac {\dot G}{G}}| \leq  4.10 \times 10^{-10} \;
yr^{-1}$. 
We also demonstrate that varying gravitational constant $G$ does not
modify cooling of white dwarfs in a significant way at least at the 
luminosities where white dwarfs are pulsationally unstable.

\end{abstract}

\section{Introduction}

There is a renewed debate in the literature 
over the issue whether the quantities known as the constants of
nature (such like G, c, h or e)  can vary with time \cite{debate}.
One of the reasons for this debate is associated with the advances
of string theory and associated ideas that the world
we live in may have more than four dimensions.

String theory \cite{string} is the only known framework which gives hopes to
reconcile gravity theory with quantum mechanics and one of its
hints is that the coupling constants appearing in low-energy
Lagrangian are determined by the vacuum expectations of some a
priori introduced massless scalar fields (dilaton or moduli
fields). The interest in physical theories with extra spatial
dimensions has also experienced considerable revival.
In the framework of multidimensional theories \cite{multidim} the four dimensional
gravity constant, for example, is in fact an effective one
appearing when integrated over additional dimensions.

Another trigger for the interest in varying the 
fundamental constants is associated with recent
observational evidence that
the fine structure constant may indeed vary with time \cite{Webb}.

All this motivates us to take seriously the possibility that (at least some of) the 
fundamental constants could
in fact vary in time.
Although from historical perspective the gravity constant $G$ was the first fundamental 
constant considered as a
dynamical quantity \cite{Dirac}, in fact it is not strongly constrained concerning its
variability
nor even measured with accuracy compatible to other fundamental constants ---
latest CODATA report raised the relative uncertainty of $G$ from $0.013 \%$ to
$0.15 \%$ \cite{Grundlach}.
In the past, there were numerous attempts to constrain the
possible temporal variability of $G$ as summarised in a recent
review paper by Uzan \cite{Uzan}.
White dwarf cooling has already been used to constrain the
temporal variation of $G$ in \cite{GarciaBerro} at the level of
white dwarf luminosity function. 

This paper is devoted to another constraint on $\frac{{\dot
G}}{G}$ obtained from asteroseismological considerations of the
ZZ Ceti pulsating star G117-B15A.
More precisely,
recent measurements of the rate of change of period of 215.2 $s$
pulsational mode \cite{Kepler2000} 
turned out to be in a very good agreement with evolutionary models 
of DAV white dwarfs comprising cooling and gravitational contraction effects.
Pulsations of DAV white dwarfs are excited as the so called g-modes which arise 
due to the competition between buoyancy and gravity forces acting on matter element displaced in a stratified medium. 
We take an advantage of this circumstance
 to derive a bound on the rate of change of the gravity constant.

\section{The role of changing $G$ in white dwarf evolution}

White dwarfs are the final stages of the evolution of
stars with initial masses of up to several $M_{\odot}$ after they ascended the
asymptotic giant branch, had lost their envelopes in the planetary nebula stage  and
developed a
degenerate carbon-oxygen cores which is an internal energy
reservoir and a thin non-degenerate envelope made up of He and H shells.
Their subsequent evolution is governed by cooling, first dominated by
neutrino losses throughout their volumes, later by surface photon emission and residual
gravitational contraction.
At certain stages of their evolution white dwarfs become pulsating stars 
(e.g. ZZ Ceti variables) and their pulsation period increases as the star cools down
\cite{Raffelt,Baglin69,jap}.

Under assumptions valid for DAV pulsating white dwarfs to which G117-B15A belongs, i.e. 
that neutrino emission is negligible and the star is in the evolutionary stage prior
to cristallization
of its core, the luminosity reads:
\begin{equation} \label{luminosity}
L = - \int_0^M T \frac{ds}{dt} \; dm =
- \int^M_0  ( \frac{du}{dt} + P \frac{d \rho^{-1}}{dt} ) \;dm
\end{equation}
where $s$, $u$, $P$ and $\rho$ have usual thermodynamical meaning with lower case letters
referring to specific (per unit mass) quantities.

The second term in (\ref{luminosity}) is in fact the work done on the star by gravitational
compression. Our goal now is to see how the assumption that $G$ is varying in time
modifies the compression term.

For this purpose let us start with hydrostatic equilibrium equation
\begin{equation} \label{hydrostatic}
\frac{dP}{dm} = - \frac{Gm}{4 \pi r^4}
\end{equation}
Multiplying (\ref{hydrostatic}) by $4 \pi r^3$ and integrating by parts one gets the
virial theorem in a standard form:
\begin{equation} \label{virial}
\Omega = - 3 \int^{M}_{0}\; \frac{P}{\rho}\;dm
\end{equation}
Differentiating gravitational energy $\Omega$ with respect to time one obtains:
\begin{equation} \label{C}
{\dot \Omega} = - 3 \int^{M}_{0}\; \frac{{\dot P}}{\rho} \; dm
+ 3 \int^{M}_{0}\; \frac{P}{\rho^2} {\dot \rho} \; dm
\end{equation}
Then the term
 $- 3 \int^{M}_{0}\; \frac{{\dot P}}{\rho} \; dm$
can be calculated by differentiating hydrostatic equilibrium equation (\ref{hydrostatic})
with respect to time, and multiplying it by
 $4 \pi r^3$ :
\begin{equation} \label{A}
4 \pi r^3 \frac{d {\dot P}}{dm} = 4 \frac{Gm}{r^2} {\dot r}
- \frac{\dot G}{G} \frac{Gm}{r}
\end{equation}

On the other hand:
\begin{equation} \label{B}
{\dot \Omega} = \int^{M}_{0} \frac{Gm}{r^2} {\dot r} \; dm + \frac{\dot G}{G} \Omega
\end{equation}

Integrating (\ref{A}) by parts over $m$ and using (\ref{B}), one gets:
$$
- 3 \int^{M}_{0}\; \frac{{\dot P}}{\rho} \; dm = 4 {\dot \Omega} - 3 \frac{\dot G}{G} \Omega
$$

Substituting this to (\ref{C}) one arrives at the formula:
$$
{\dot \Omega} = \frac{\dot G}{G} \Omega - \int^{M}_{0}\; \frac{P}{\rho^2} {\dot \rho} \; dm
$$
This means that the compression term in (\ref{luminosity}) is equal to
\begin{equation} \label{compression}
\int^M_0 P \frac{d \rho^{-1}}{dt} \;dm = {\dot \Omega} - \frac{\dot G}{G} \Omega
\end{equation}

Hence the luminosity of a white dwarf reads
\begin{equation} \label{lumG}
L =  - \frac{d E}{dt} + \frac{\dot G}{G} \Omega
\end{equation}
where: $E = U + \Omega$ denotes total energy of the star and respective formula 
in Garcia-Berro
et al. \cite{GarciaBerro} is recovered.

However this does not mean that varying G can influence white dwarf cooling in a 
significant way (at least during most of their evolution).
In order to see this, let us consider total internal energy of the star
$U$ as a function of state $U(\rho, T)$
$$
\frac{d U}{dt} = \int_0^M \frac{d u}{dt} \; dm =
\int_0^M (\frac{\partial u}{\partial T} \frac{d T}{dt} +
 \frac{\partial u}{\partial \rho} \frac{d \rho}{dt}) \; dm
$$
using well known identity:
$ - \rho^2 \frac{\partial u}{\partial \rho} = T \frac{\partial P}{\partial T} - P $
one gets:
$$
\frac{d U}{dt} = \int_0^M \frac{\partial u}{\partial T} \frac{d T}{dt} \; dm + 
\int_0^M T \frac{\partial P}{\partial T} \frac{d \rho^{-1}}{dt}  \; dm
- \int_0^M P \frac{d \rho^{-1}}{dt}  \; dm  = \frac{d E_{thermal}}{dt} +
\frac{d E_{grav}}{dt} - \frac{d \Omega}{dt}
+ \frac{\dot G}{G} \Omega
$$
rearranging this we obtain the following:
$$
\frac{d E}{dt} = \frac{d U}{dt} + \frac{d \Omega}{dt} = \frac{d E_{thermal}}{dt} +
\frac{d E_{grav}}{dt}
+ \frac{\dot G}{G} \Omega
 $$
After substituting this result to (\ref{lumG}) one can see that $\dot G$ terms
will cancel leading to the standard formula:

$$
L =  - \frac{d E_{thermal}}{dt} - \frac{d E_{grav}}{dt}
$$

Here, as commonly accepted, $E_{thermal}$ denotes thermal energy of the star, and $E_{grav}$ - the fraction
of gravitational energy $\Omega$ contributing to the luminosity.

Although formula (\ref{lumG}) is formally correct, yet the
cancellation of $\dot G$ terms is still present in it in a hidden
form. For this reason, basing on a standard lore, one is
tempted (as we indeed were) to identify the change in the total
energy of the star $\frac{d E}{dt}$ with the change of its thermal
energy $\frac{d E_{thermal}}{dt}$ without noticing that the part
of $\frac{d E}{dt}$ coming from $\dot G$ gets cancelled with
$\frac{\dot G}{G} \Omega$ term in (\ref{luminosity}). 

Because $\frac{d E_{thermal}}{dt} = \int^M_0 c_v \frac{d T}{dt} \; dm$ and
$\frac{d E_{grav}}{dt} = \int^M_0 T \frac{\partial P}{\partial T} \frac{d \rho^{-1}}{dt}  \; dm$ 
one can see immediately that (even with changing $G$)
zero temperature "black dwarf" configuration would have zero luminosity as expected.

In order to assess the real contribution of varying $G$ to the white dwarf cooling 
let us assume that we have highly degenerate electron gas, ideal gas of ions and Coulomb
correction: $P = P_e + P_i + P_{C}$, $P_e = \frac{2}{5} Z \psi P_i$, $P_C = -0.3\Gamma P_i$ where $\psi$ is degeneracy
parameter and as usually $\Gamma$ measures Coulomb potential energy compared with thermal energy.
Then to the first order in $1 / \psi$
\begin{eqnarray}
 \frac{d E_{grav}}{dt} &=& \int^M_0 P_i \frac{d \rho^{-1}}{dt}  \; dm
= \int^M_0 \frac{P_i}{P_i+P_C+P_e} P \frac{d \rho^{-1}}{dt}  \; dm = \\
 &=& \int^M_0 \frac{P}{1 - 0.3 \Gamma + 0.4 Z \psi}  \frac{d \rho^{-1}}{dt}  \; dm = \\
&=& 
\langle (1 - 0.3 \Gamma + 0.4 Z \psi)^{-1} \rangle ({\dot \Omega} - \frac{\dot G}{G} \Omega)
\end{eqnarray}
where the average above is defined by equation. Now, in carbon-oxygen white dwarfs like G117-B15A
typically $\psi \approx 550$ and $\Gamma \approx 55$. Hence the modification to star's luminosity coming from
varying $G$ is about three orders of magnitude smaller than in the case of naive interpretation of formula (\ref{lumG}). At low luminosities, however, close to the luminosity function cut-off the effect of varying $G$ could be non-negligible as pointed out in \cite{GarciaBerro}. However, in light of both theoretical and observational uncertainties in precise determination of the cut-off the accuracy of luminosity function method is not a great one.

Although changing $G$ has no direct effect on cooling properties of white dwarfs during most of their evolution there hopefully exists certain possibility to obtain such a bound from DAV pulsating white dwarfs. The basic idea behind is simple: because DAV oscillations are driven by gravity force (acting against bouyancy force) the resulting oscillation period can be used to "measure" $G$ in a similar manner like in an elementary experiment one measures the acceleration of gravity $g$ from measuring the period of a pendulum.  

Extensive theoretical studies of non-radial oscillations of white
dwarfs had been carried out long time ago \cite{Baglin69}.
Appropriate theory of stellar oscillations consists in linearizing
the Poisson equation as well as equations of momentum, energy and
mass conservation with respect to small non-radial perturbations \cite{Unno,Cox}.

DAV pulsating white dwarfs oscillate in non-radial g-modes whose main 
restoring force is gravity (acting against buoyancy force). 
In spherically symmetric stars, g-modes can 
be represented as standing waves 
$f'_{klm}(r, \theta, \phi, t) = g'_{klm}(r) Y^m_l(\theta, \phi) 
\exp( - i \sigma_{klm} t)$ where by $f'$ we denoted Eulerian perturbation of 
given quantity e.g. the  pressure or density. G-modes observed in DAV pulsators have modest radial 
orders $1 \leq k \leq 25$ and low angular degrees $1 \leq l \leq2$ (the latter 
is probably an observational selection effect).  
Because buoyancy is the restoring force for g-modes, the 
Brunt-V$\rm {\ddot a}$is$\rm {\ddot a}$l$\rm {\ddot a}$ 
frequency $N$ is the most important quantity setting the scale in the pulsation spectrum. 
Indeed in the asymptotic theory, eigen frequencies of g-modes are proportional to $N$. The Brunt-V$\rm {\ddot a}$is$\rm {\ddot a}$l$\rm {\ddot a}$ 
frequency is usually expressed as:
\begin{equation} \label{BVfreq}
N^2 = - g A = - g ( \frac{d ln \rho}{d r} - \frac{1}{\Gamma_1} \frac{d ln p}{dr})
\end{equation}
where: $g$ denotes local gravity, 
$\rho$ - density, $r$ - radial coordinate, $p$ is the pressure and 
$\Gamma_1$ is the adiabatic index \cite{Cox}. In the context of white 
dwarfs alternative expression given by Brassard is even more useful
\begin{equation} \label{Brassard}
N^2 = \frac{g^2 \rho}{p}\; \frac{\chi_T}{\chi_{\rho}} (\nabla_{ad} - \nabla + B)
\end{equation}
where $\chi_a := \frac{\partial ln p}{\partial a}$ $a = T, \rho$ and 
$B = - \frac{1}{\chi_T} \sum^{N-1}_{i=1} \chi_{X_i} \frac{d ln X_i}{d ln p}$ 
with $X_i$ denoting the mass fraction of atomic species $i$. 
From the above formula one can see that the C/O, He/C and H/He transition zones
manifest themselves as three peaks in $N^2$ which play an important role in 
modifying the period structure leading to the phenomena of mode trapping (near the surface) 
or mode confining (at central parts of the star) \cite{Bradley96}. 
In fact similar asteroseismological considerations have already been used to calibrate 
compositional stratification of G117-B15A star \cite{Salaris,Corsico,Bradley96}. 

For a
zero-temperature degenerate electron gas $A=0$ and hence $\sigma_{klm} = 0$ meaning that
no g-modes are supported. However, if non-zero thermal effects are
taken into account one can show \cite{Baglin69} that $A \propto T^2$ and consequently $\sigma_{klm}^2 \propto T^2$ i.e. the
periods scale like $1/T$ where $T$ is the temperature of
isothermal core. For high overtone modes one has:
\begin{equation}
P^2 \approx - \frac{k^2 r^2}{A g l(l+1)} 
\end{equation}

Following Winget, Hansen and Van Horn \cite{WH83} (see also \cite{Kepler2000,other}) 
one can write:
\begin{equation} \label{Pdot}
\frac{d ln P}{dt} = - a \frac{d ln T}{dt} + b \frac{d ln R}{dt}  
\end{equation}
where: coefficients $a$ and $b$ are positive numbers of the order of unity. 
Since our further calculations will be an order-of-magnitude estimates we shall 
assume the $a$ and $b$ coefficients as equal to 1. 
First term comes from white dwarf cooling and dominates over the term 
from gravitational contraction.  
In fact DA white dwarf models (representative for ZZ Ceti pulsators such like G117-B15A) 
give a core temperature of $1.2 \times 10^7 K$ and a cooling rate of $0.005 K$ \cite{Kepler2000}. 

In the case of varying $G$ formula (\ref{Pdot}) gets modified 
\begin{equation} \label{Pdot}
\frac{d ln P}{dt} = - a \frac{d ln T}{dt} + b (\frac{d ln R}{dt} - \frac{d ln G}{dt}) 
\end{equation}
This means that time varying $G$ affects the contraction term which is in agreement with underlying physics as described above. 

\section{Asteroseismology of G117-B15A and bound on variability of $G$}

Since its discovery in 1976 \cite{McGraw} G117-B15A has been extensively studied and its
basic physical properties are quite well established.
Regarding its variability the observed periods are 215.2, 271 and 304.4 s together with higher
harmonics and linear combinations thereof \cite{Kepler82}.
Quite recently, with a considerable time interval of acquired data, Kepler et al.
\cite{Kepler2000} recalculated the
rate of period increase and found a value of
${\dot P} = (2.3 \pm 1.4) \times 10^{-15} \;\;s\;s^{-1}$.
It has been claimed \cite{Corsico} that the 215.2 s mode of G117-B15A is the
most stable oscillation ever recorded in the optical band (with a stability compared to
milisecond pulsars) and that observed ${\dot{P}}$ value is really due to evolutionary effects.
This circumstance has already made it possible to derive bounds on axion mass \cite{Corsico}
and the compactification scale $M_s$ in the Arkani-Hamed--Dimopuolos--Dvali
theory with large extra dimensions \cite{BiesiadaMalec}.

The white dwarf pulsator G117-B15A has a mass of
$0.59 \; M_{\odot}$ (established spectroscopically), 
effective temperature $T_{eff} = 11 620\;K$ \cite{Bergeron}
 and luminosity
$log ({\cal L}/{\cal L}_{\odot}) = -2.8$ \cite{McCook}
(i.e. ${\cal L} = 6.18 \times 10^{30}\;\;erg \;s^{-1}$).
More recent study by Koester and Allard \cite{KA} suggested the lower value 
for the mass $0.53 \; M_{\odot}$. 
Typical model for such CO star (with C:O = 20:80 \cite{Bradley96} or C:O=17:83 \cite{Salaris})
predicts the central
temperature $T = 1.2 \times 10^7\;\;K$ \cite{Corsico}
and the radius $R = 9.6 \times 10^8 \;\; cm$ \cite{Kepler2000}. 
Knowledge of the internal chemical profile of the core is essential for the following reasons. First, the heat capacity 
of the core is mainly due non-degenerate ions and is inversely proportional to the average atomic number, thus enhanced O 
abundance implies less heat capacity, faster cooling and hence larger $\dot P$. Second, the smaller C abundance 
implies steeper abundance slopes at the outer boundary of the core. This can lead to different radial configurations of 
the modes. It turns out that the 215.2 s mode has large amplitude in this interphase and can be used to constrain the 
admissible range of C abundance \cite{Corsico}. 
The chemical composition of the white dwarf interior as a function of mass 
calculated by Salaris et al. \cite{Salaris} are in agreement with Bradley's 
results concerning DA pulsators \cite{Bradley96} as well as with the work of Dominguez et al. who computed 
updated chemical profiles for the AGB progenitors of white dwarfs finding similar stratifications \cite{Dominguez}. 
Therefore they can serve as a reliable reference. 

There are two main processes which 
govern the rate of period change in
theoretical models of ZZ Ceti stars: the cooling of the star (which increases the period
as a result of increasing degeneracy) and residual gravitational contraction (which shortens the
period) \cite{Baglin69}. For the G117-B15A star the contraction rate is negligibly
small \cite{Kepler2000}
hence confirming the assumption that cooling
is a dominant process for this star. Theoretical value of $\dot P$ used below in
our estimates
was derived
from numerical calculations \cite{Corsico} in which such issues as gravitational contraction,
chemical composition of the star or properties of outer He-H layers were carefully
taken into account. This
value is also in agreement with previous independent evolutionary calculations \cite{Bradley98}.
Therefore, the correction from residual gravitational contraction is already present in
theoretical rate of period change adopted here. 

The difference between observed (labeled "O") and calculated (labeled "C")
logarithmic time derivatives of the period 
(precisely an absolute value thereof)
$|\Delta(\frac{\dot P}{P})| := |(\frac{\dot P}{P})_O-(\frac{\dot P}{P})_C|$ reads:
$|\Delta(\frac{\dot P}{P})|=|{\frac {\dot G}{G}}|$.
So the upper bound for $|\Delta(\frac{\dot P}{P})|$ translates into a bound on the rate of change
of $G$.

The most recent determination \cite{Kepler2000} of $\dot P$ for the $215.2 \;s$ mode gives
the value $(2.3 \pm 1.4) \times 10^{-15} \;\;s\;s^{-1}$ whereas
theoretical prediction for the rate of change of the period is
$3.9 \times 10^{-15} \;\;s\;s^{-1}$ \cite{Corsico}. 
A conservative method of constraining the admissible role of varying $G$ in the discrepancy 
between theory and observations comes from 
the fact that the theoretical value falls within a two-sigma interval for the
observed value. Therefore current observational knowledge concerning G117-B15A pulsating
star and associated consequences for the $\dot P$ of the $215.2 \;s$ mode can be translated
into the following
bound for the rate of change of the gravitational constant:
\begin{equation} \label{bound}
|{\frac {\dot G}{G}}| \leq 4.10 \times 10^{-10} \;
yr^{-1}
\end{equation}

Of course, while constraining the variability of $G$ in a manner described above 
we have neglected other non-standard effects. For instance a larger 
(than currently thought admissible) 
ammount of axion \cite{Corsico} or Kaluza-Klein graviton \cite{BiesiadaMalec} emission  
could in principle 
be compensated by respective variation of $G$. Not having in mind any specific theory of 
varying $G$ (nor any other exotic physical theory) we disregarded the issue of joint 
effect of varying $G$ and extra cooling by emission of exotic particles.  
Of other effects influencing the mode properties in a way that may interfere with changing $G$,  
the beginning of cristallization would slow down the cooling rate allowing for larger $\dot G$. 
However, it has been argued \cite{Winget97} that G117-B15A is not cool enough to have a crystallized 
core.

\section{Conclusions}

It would be instructive to compare the derived bound on the rate of change of the
gravitational constant with already existing bounds. They can be roughly divided into
four classes: geophysical bounds, bounds from celestial mechanics, astrophysical and
cosmological bounds. The extensive discussion of these issues can be found in a review
paper by Uzan \cite{Uzan}. Let us here invoke only the strongest (or the most representative)
bounds from these four categories and discuss their relation to our bound.

In the class of geophysical/paleonotological bounds the strongest one is the bound
$-\frac{\dot G}{G} \leq 8 \times 10^{-12}\;\;yr^{-1}$ coming from the stability of radii
of the Earth, the Moon and Mars as estimated by McElhinny \cite{McElhinny}. Celestial
mechanics arguments in the Solar System refer either to systematic deviations from Keplerian
orbital periods in the case of time varying $G$ (which can reveal themselves e.g. in
lunar occultations)
or
to monitoring of the separation of orbiting bodies (lunar laser ranging or planetary ranging).
The most stringent bounds obtained by these techniques are:
$\frac{\dot G}{G} \leq (3.2 \pm 1.1) \times 10^{-11}\;\;yr^{-1}$ \cite{vanFlandern81}
and
$\frac{\dot G}{G} \leq (-2. \pm 10.) \times 10^{-12}\;\;yr^{-1}$ \cite{Shapiro90}, respectively.

In the case of
pulsar timing data Kaspi et al. \cite{Kaspi94} derived the strongest bounds in this
class for PSR B1913+16 as:
$\frac{\dot G}{G} \leq (4. \pm 5.) \times 10^{-12}\;\;yr^{-1}$.

Cosmological bounds come out of the big bang nucleosynthesis (BBN)
considerations and the most stringent ones are those of Accetta et
al. \cite{Accetta90} ($|\frac{\dot G}{G}| \leq 9. \times
10^{-13}\;\;yr^{-1}$) and Rothman and Matzner \cite{RM82}
($|\frac{\dot G}{G}| \leq 1.7 \times 10^{-13}\;\;yr^{-1}$). They
are however valid only for the Brans-Dicke theory. Moreover, the
BBN arguments based on the abundance of light elements are
sensitive not only to the gravity constant but also (and to
greater extent) to fine structure constant and other factors like
the neutron to proton ratio, the neutron-proton mass difference or
the number of neutrino families.

The most stringent astrophysical constraint
$|\frac{\dot G}{G}| \leq 1.6 \times 10^{-12}\;\;yr^{-1}$ comes from helioseismology
\cite{Guenther98}. However it has been derived for a specific Brans-Dicke type theory 
with varying $G$.
Previous white dwarf constraint based on a position of the
cut-off on the luminosity function of white dwarfs \cite{GarciaBerro} was
$-\frac{\dot G}{G} \leq 3.^{+1}_{-3} \times 10^{-11}\;\;yr^{-1}$. 

In conclusion one can say that
the bound derived here for ${\dot G}$ from the observed
rate of change of the pulsational period of G117-B15A supports the claims
first expressed in \cite{Corsico} that this star is becoming a tool for fundamental
physics. Moreover, there are other DAV pulsating stars such like L19-2 or R 548 
\cite{Isern93} for which the rate of change of the period can be measured in forthcoming 
WET campaigns. Having high precision asteroseismological data for a larger set of objects 
one would be able to test exotic physical theories (such like those in which $G$ varies) 
with much better accuracy.

\section{Acknowledgements}
The authors are grateful to Bohdan Paczy{\'n}ski for a stimulating discussion on the white dwarf cooling with varying $G$ and to the referee for clarifying remarks on the luminosity function method.

\end{document}